\documentclass[aip,jcp,preprint,floatfix]{revtex4-1}
\usepackage[dvips]{graphicx}
\usepackage[dvips]{epsfig}
\usepackage{amsmath}
\newcommand{\beq}{\begin{equation}}
\newcommand{\eeq}{\end{equation}}

\newcommand{\HSb}{\mbox{H$_{3}$Sb$_{3}$P$_{2}$O$_{14}$}}
\newcommand{\un}[1]{\ensuremath{\unskip\,\mathrm{#1}}}
\DeclareGraphicsExtensions{.eps}
\usepackage{hyperref}

\begin{document}

\title{Slow dynamics of a colloidal lamellar phase}
\author{Doru Constantin}
\email{constantin@lps.u-psud.fr}
\author{Patrick Davidson}
\affiliation{Laboratoire de Physique des Solides, Universit\'{e}
Paris-Sud, CNRS, UMR 8502, 91405 Orsay, France.}

\author{\'{E}ric Freyssingeas}
\affiliation{Laboratoire de Physique, \'Ecole Normale Sup\'erieure
de Lyon, CNRS, UMR 5672, 69364 Lyon, France.}

\author{Anders Madsen}
\affiliation{European Synchrotron Radiation Facility, Bo\^{\i}te
Postale 220, 38043 Grenoble, France.}

\date{\today}

\begin{abstract}
We used X-ray photon correlation spectroscopy to study the dynamics
in the lamellar phase of a platelet suspension, as a function of the particle concentration. We measured the collective diffusion coefficient along the director of the phase, over length scales down to the inter-particle distance, and quantified the hydrodynamic interaction between the particles. This interaction sets in with increasing concentration and can be described qualitatively by a simplified model. No change in the microscopic structure or dynamics is observed at the transition between the fluid and the gel-like lamellar phases.
\end{abstract}



\maketitle

\section{Introduction}

In order to reach a thorough understanding of colloidal suspensions,
it is crucial to probe their dynamics at length scales comparable to
the inter-particle distance. Thus, a considerable amount of experimental and
theoretical work has been concerned with the dynamics of dense
colloidal suspensions\cite{Pusey89}. Space- and time-resolved experiments were mainly performed using dynamic light scattering (DLS) on micron-size spherical particles, both in the isotropic\cite{Segre95} and the crystalline\cite{Cotter87} phases.

DLS was also used to study dilute isotropic suspensions of elongated particles \cite{Graf91} or solutions of disk-shaped particles close to the sol-gel transition \cite{Kroon96}, but using this technique for the study of concentrated and/or ordered solutions becomes challenging, due to the required particle sizes. This is nevertheless a regime where the system should exhibit rich dynamics, influenced by the (short- or long-range) order and by the hydrodynamic coupling, the latter being quantified by the \textit{hydrodynamic function}.

Theoretical efforts aiming to calculate this function have focussed mainly on suspensions of spherical particles \cite{Naegele97}; they have been largely validated by the experimental results (see below). For anisotropic particles, on the other hand, the calculations are much more involved and they were mostly restricted to solutions of slender rods \cite{Batchelor71,Shaqfeh89}. We are not aware of any analytical results for the hydrodynamic function of plate-like particles. Experimental measurements at high scattering vector (corresponding to the interparticle distance) are also lacking, due to the intrinsic  wavelength limitation of DLS.

In recent years, the accessible distance range increased significantly via the use of X-ray Photon Correlation Spectroscopy (XPCS), which is fundamentally similar to DLS but uses X-ray radiation as a probe. Although this technique is only effective on systems with high scattering contrast and slow relaxation rates, it has already been used to study concentrated hard sphere solutions \cite{Robert05,Banchio06,Robert08}, aging suspensions \cite{Bandyopadhyay04,Wandersman08}, particles dispersed in complex fluids \cite{Caronna08,Guo09}, and interface dynamics \cite{Gutt03,Gutt07}.

Some XPCS studies were performed on mesophases. For instance, the collective diffusion coefficient of a colloidal nematic phase was determined over a wide range of the wave vector $q$. The analysis showed that the dynamics of the system displays strong slowing down over length scales larger than the interparticle distance \cite{Constantin10}. The relaxation of capillary surface waves has also been measured \cite{Madsen03}. Smectic phases --which have a certain degree of positional order-- represent ideal candidates for XPCS investigations, as the scattering intensity is concentrated in the vicinity of the Bragg peaks, leading to high count rates. That is why some of the first XPCS results were obtained on such systems, in particular under the form of free-standing \cite{Price99,Sikharulidze02,Sikharulidze03} films. The bulk dynamics of a lamellar lyotropic phase of surfactant were also studied \cite{Constantin06}. The results, confirmed by DLS measurements, could be interpreted in terms of the continuum elastic theory of smectics \cite{Martin72,Ribotta74}.

In this paper we present results obtained on lamellar lyotropic phases composed of large inorganic colloids; in this system, the slow relaxation rates and the high scattering contrast greatly extend the accessible $q$-range. We measure the collective diffusion coefficient $D(q_z)$, where $\hat{z}$ is the direction of the lamellar director, for a fairly wide range of scattering vectors along the lamellar director, $q_z$ (covering at least the first Bragg peak of the phase). We then obtain the hydrodynamic function $H(q_z)$. The large aspect ratio of the particles and their lamellar order enable us to describe the dynamics using a simplified analytical model, which is in semi-quantitative agreement with the data. As a function of the concentration, the hydrodynamic coupling goes from very weak to extremely strong.

\section{Experimental}
\label{sec:exp}

We used concentrated suspensions of phosphatoantimonate (\HSb) particles, with a typical thickness of 1~nm and at least 300~nm wide \cite{Gabriel01}, synthesized as in reference \onlinecite{Piffard86}. The lateral size distribution of the particles was assessed using a scanning electron microscope equipped with a field emission gun (SEM-FEG). In a typical experiment, a colloidal suspension was diluted down to a volume fraction of $2.5 \, 10^{-4}$ by adding distilled water. Then, a drop of suspension was spin-coated at $\sim 1000$~rpm onto a silicon wafer. Close inspection of the samples, by SEM-FEG at 10~kV and by optical microscopy, revealed the presence of plate-like particles that sometimes had a clear hexagonal shape. Their size distribution is extremely broad and ranges from a few hundred nanometers to more than five microns.

The batch solutions were diluted, either with pure water or with a 50:50 (wt\%) glycerol/water mixture to reach the desired volume concentration $\phi$. The room-temperature phase diagram was determined by visual inspection of the vials, in natural light and between crossed polarizers. The birefringent lamellar phase has a gel-like texture above $\phi = 1.8 \%$ (does not flow under its own weight, i.e.\ it has a zero-frequency elastic modulus) and is fluid below this value, down to $\phi = 0.65 \%$ where it coexists with a very dilute isotropic phase, with a concentration $\phi _I < 0.1 \%$. These phase boundaries are compatible with the phase diagram presented in Figure 3 of reference~\onlinecite{Gabriel01}, for the case of no added salt (ionic strength below $10^{-4}$~mol/l).

The lamellar samples were prepared by aspiration in
round glass capillaries, 1~mm wide and with a wall thickness of 10~$\mu$m. By scanning the capillaries in the beam, it is easy to find aligned domains (usually, with the director normal to the capillary walls).

The small-angle X-ray scattering (SAXS) and XPCS measurements were
performed at the TROIKA beam line ID10A of the ESRF with an X-ray
energy of 8~keV ($\lambda=1.55$~\AA) selected by a single-bounce
Si(111) monochromator, in the uniform filling mode of the storage
ring. A (partially) coherent beam is obtained by inserting a 10
$\mu$m pinhole aperture a few centimeters upstream of the sample.

\begin{figure}[htbp]
\centerline{\includegraphics[width=8.4cm,
keepaspectratio=true]{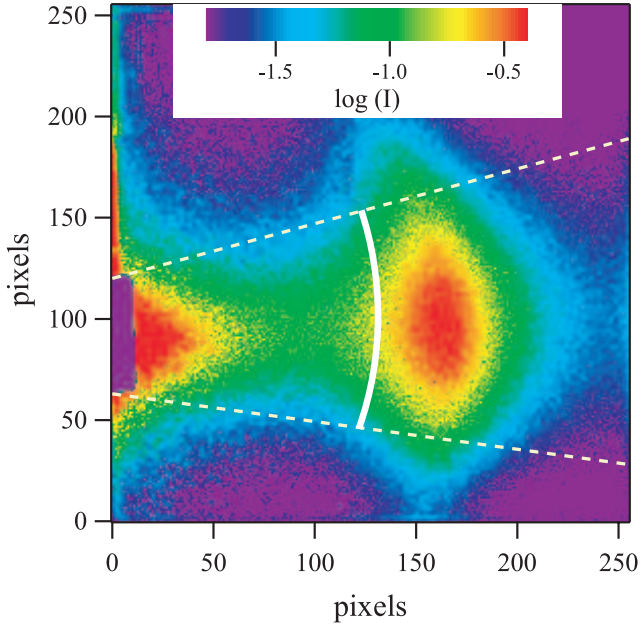}} \caption{SAXS scattering pattern for
the sample with $\phi = 2.1$~vol\%; the lamellar director is roughly
horizontal. The feature at the center of the image is the first Bragg
sheet. The shadow of the beamstop is visible at the left edge. The
averaging was done within the range delimited by the two dashed
lines, on circular arcs corresponding to a given scattering vector
$q$. One arc is shown as solid line, for $q = 14.5 \, 10^{-3} \un{\AA}^{-1}$.} \label{Fig:Static}
\end{figure}

We used a 2D Maxipix detector consisting
of $256 \times 256$ square pixels (55~$\mu$m in size) and with acquisition rates of up to 1~kHz. For XPCS, a few thousand images were recorded and the
intensity autocorrelation functions were calculated by ensemble
averaging \cite{Fluerasu07} over pixels with the same absolute value
of the scattering vector (see Figure \ref{Fig:Static}). The SAXS pattern is obtained as the average intensity over all the images in each run. Since we are only concerned with the $q_z$ dependence of the measured quantities, in the following we will denote throughout $q_z$ by $q$.

\begin{figure}[htbp]
\centerline{\includegraphics[width=8.4cm,
keepaspectratio=true]{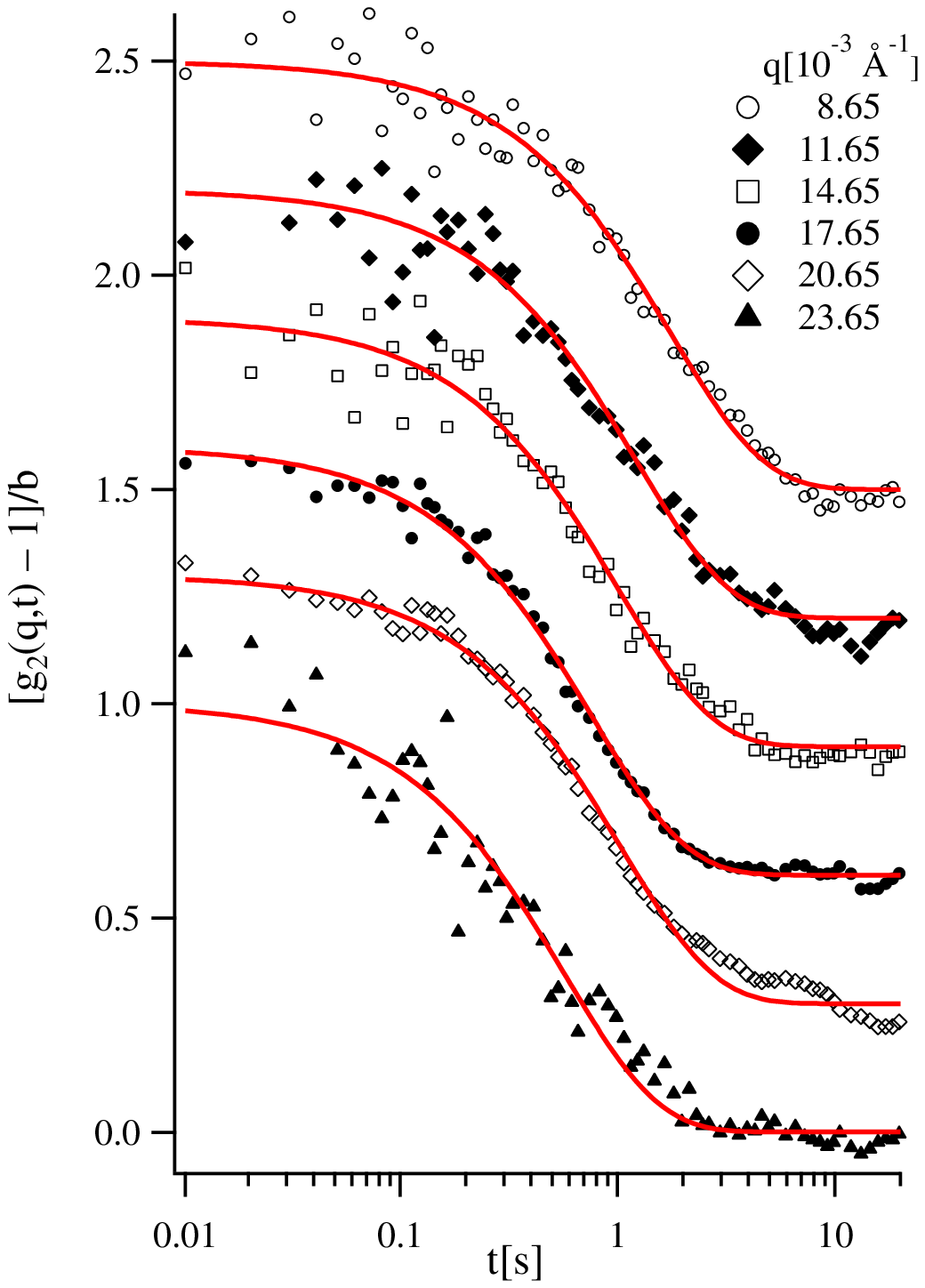}} \caption{Normalized intensity
correlation function $g_2 (q,t)$ for the sample with $\phi =
2.1$~vol\% (see Figure \ref{Fig:Static}), shifted vertically in
steps of 0.3. The different symbols correspond to different values
of the scattering vector $q$. The solid lines are fits to Equation
(\ref{eq:g2qt}).} \label{Fig:g2qt}
\end{figure}

The relaxation of concentration fluctuations with a given wave
vector $q$ is reflected in the field correlation function at that
value of the scattering vector: $g_1(q,t)=\left \langle
E^*(0)E(t)\right \rangle / \left \langle E^*(0)E(0)\right \rangle$.
As we will see below, our data is well described by a single
exponential relaxation: $g_1(q,t)=\exp [-\Omega(q) t]$.

\begin{figure}[htbp]
\centerline{\includegraphics[width=8.4cm,
keepaspectratio=true]{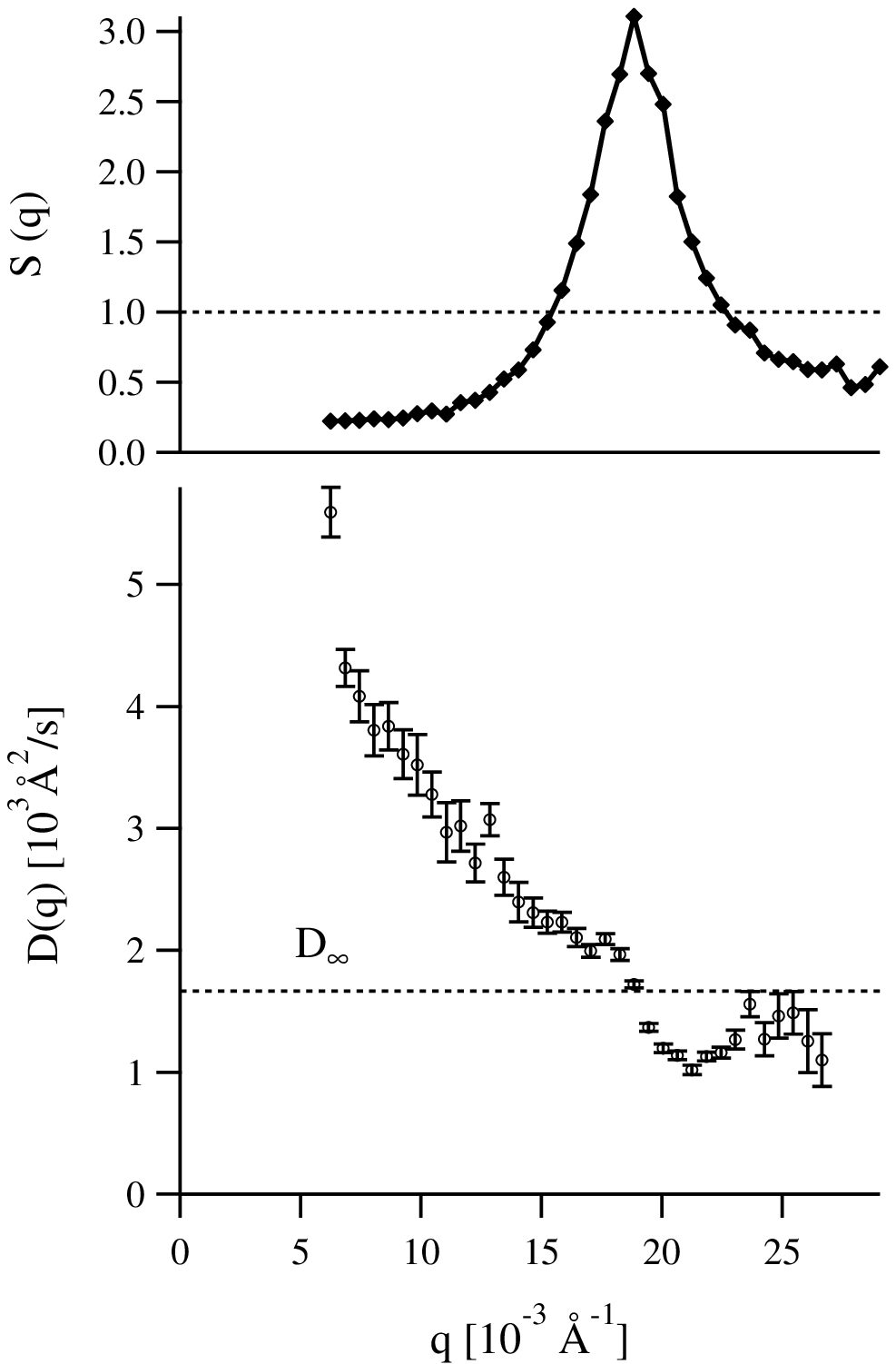}} \caption{Structure factor $S(q)$ and
diffusion constant $D(q)$ for the sample with $\phi = 2.1$~vol\%.
$D_{\infty}$ (materialized by a dotted line) is the high-$q$ value of the
diffusion constant, corresponding to the range where the structure
factor saturates to 1.} \label{Fig:Dynamic}
\end{figure}

\begin{figure}[htbp]
\centerline{\includegraphics[width=8.4cm,
keepaspectratio=true]{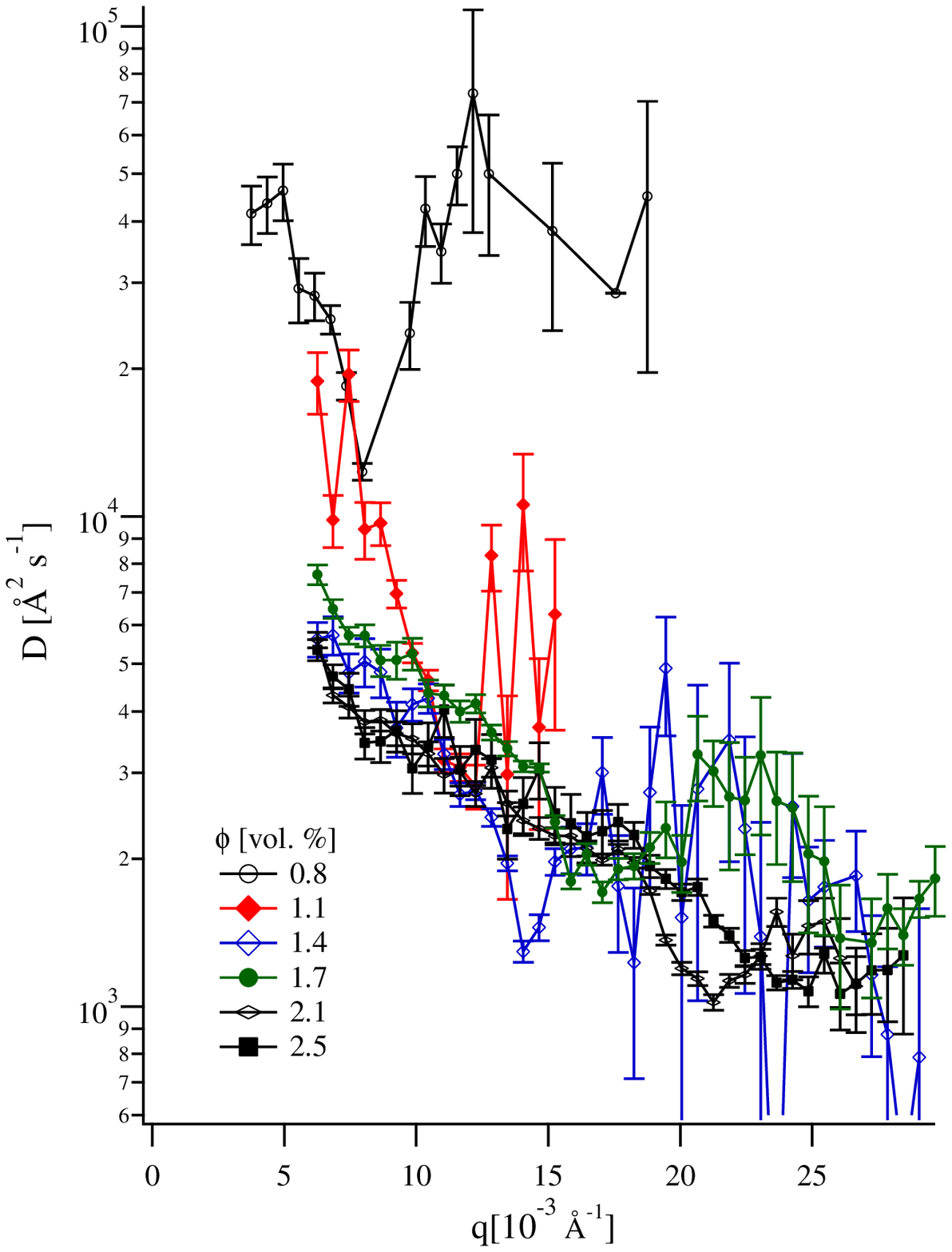}} \caption{Scale-dependent diffusion constant $D(q)$ for different $\phi$ values. The ``dips'' visible in the curves with $\phi = 0.8$ and 1.1~vol\% (and, to a lesser degree, for $\phi = 1.4$~vol\%) occur at the position of the peak in the static structure factor $S(q)$.} \label{Fig:Dall}
\end{figure}

Experimentally, we measure the normalized intensity correlation
function $g_2(q,t)$, related to the field correlation function by
the Siegert relation:
\begin{equation}
\label{eq:g2qt} g_2(q,t)=1 + b(q) \left | g_1(q,t)\right |^2 = 1
+ b(q) \exp [-2 \Omega(q) t]
\end{equation}
where the contrast factor $b$ is a few percent. Fitting the
experimental data to Equation (\ref{eq:g2qt}), as illustrated in
Figure \ref{Fig:g2qt}, yields the relaxation rates $\Omega(q)$. We
will further define a scale-dependent diffusion constant, $D(q) =
\Omega(q) / q^2$. In the following, we will work with the (static)
structure factor $S(q)$ and with the diffusion constant $D(q)$. For
the sample with $\phi = 2.1$~vol\% (used as an example in Figures \ref{Fig:Static}, \ref{Fig:g2qt} and \ref{Fig:Dynamic}) these parameters are displayed in Figure
\ref{Fig:Dynamic}. The diffusion constant is shown for all $\phi$ values in Figure \ref{Fig:Dall}.

\section{Results and Discussion}

The structure factor and fluctuations of the smectic phase have been studied in detail over the last decades by many authors (see Ref. \onlinecite{deJeu03} for an in-depth review). The deformation free energy of the phase consists essentially of two terms, proportional to the compression modulus (along the smectic director) and to the bending stiffness of the layers. Both the amplitude and the relaxation of the deformation eigenmodes (corresponding to a unique phonon) can be described in a relatively straightforward (although algebraically cumbersome) fashion.

However, if one is interested in the dynamic structure factor, i.e.\ in the collective behavior of the system at a given wave vector, the calculations become extremely involved, since all the phonons contribute in non-trivial ways. In order to make the analytical treatment as easy as possible, so that the underlying physics is not obscured by the mathematical formalism, some simplifying assumptions must be made.

In this work, we are dealing with rather dilute phases composed of very stiff and very large platelets. We will therefore assume that the compression modulus is low and the bending stiffness is large, so that the most important fluctuations are those affecting the spacing between platelets along the director $\hat{z}$ and we treat the system as effectively one-dimensional, within the framework of a damped harmonic chain model.

\subsection{Statics}

For this model, the static structure factor has an analytical solution, given by Refs. \onlinecite{Emery78,Radons83}:
\begin{equation}
I(q,0) = S(q) = \frac{\sinh \left ( \frac{q^2 \sigma ^2}{2} \right )}
{\cosh \left ( \frac{q^2 \sigma ^2}{2} \right ) - \cos \left ( q d \right )}
\label{Sq}
\end{equation}
\noindent with $\sigma$ the typical fluctuation amplitude, defined
by $\sigma ^2 =k_BT/\alpha$ as a function of the strength of the
harmonic potential, $\alpha$.

\begin{figure}[htbp]
\centerline{\includegraphics[width=8.4cm,
keepaspectratio=true]{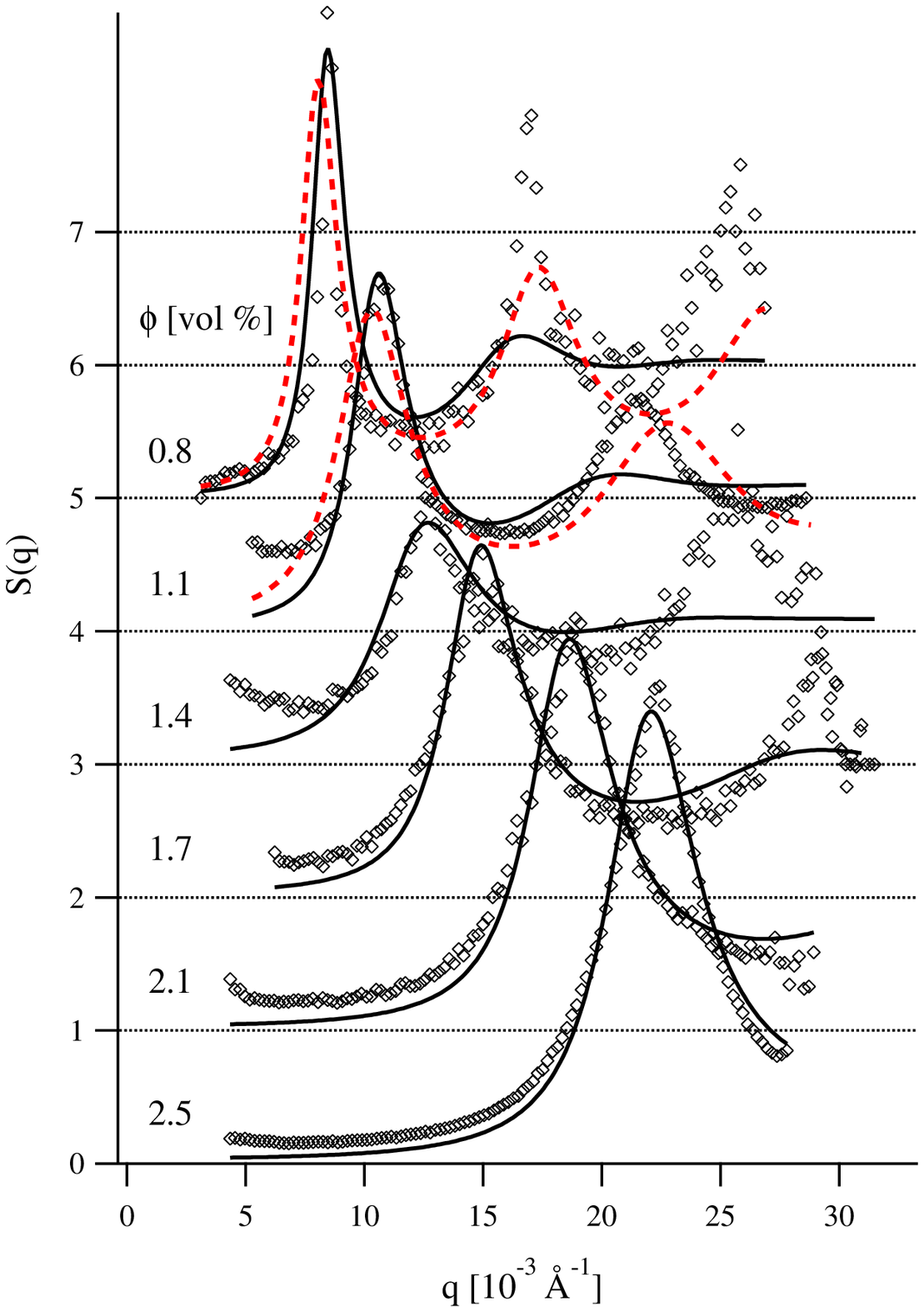}} \caption{Structure factor
$S(q)$ for different concentrations $\phi$ of the lamellar phase
(diamonds). Curves shifted vertically in steps of 1. The solid lines
are fits to the harmonic chain model (\ref{Sq}). For the two lowest
concentrations, we also present as dashed lines the fit with the
hard-rod model (see the text for more details).} \label{Fig:Sq100Ia}
\end{figure}

Clearly, the harmonic model is not very accurate at high $q$, as one can see in Figure \ref{Fig:Sq100Ia} for the fits to the low-concentration data. The interaction potential is probably stiffer, leading to pronounced second- and even third-order peaks, which are better described by a hard-particle model \cite{Wertheim64} (plotted as dashed lines). In the following we will nevertheless use the harmonic approximation, which describes the first peak rather well and which is more tractable as far as the dynamics are concerned (see Section \ref{sec:dyn}).

\begin{figure}[htbp]
\centerline{\includegraphics[width=8.4cm,
keepaspectratio=true]{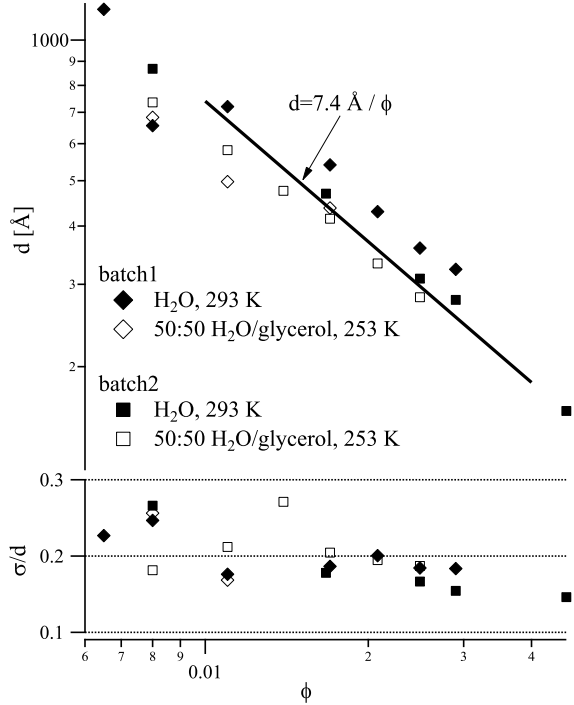}} \caption{Top: dilution law.
The smectic repeat distance $d$ is plotted against the volume
fraction $\phi$ in log-log representation. The solid line
represents the $d \sim \phi^{-1}$ variation expected for a
one-dimensional system. The resulting platelet thickness, $7.4 \pm 1.2 \, \un{\AA}$, is clearly below the 10~{\AA} value given in the
literature \cite{Gabriel01}. Bottom: the ratio between the
fluctuation amplitude $\sigma$ (see Equation \ref{Sq}) and $d$. The
data is shown for all the available samples, both in pure water at
293~K and in a 50:50 glycerol/water mixture at 253~K.}
\label{Fig:dilution}
\end{figure}

From the analysis of the structure factor data in Figure \ref{Fig:Sq100Ia} using the model (\ref{Sq}) we extract the repeat distance $d$ and the fluctuation amplitude $\sigma$. In Figure \ref{Fig:dilution} we present both the dilution law $d(\phi)$ and the ratio $\sigma /d$ as a function of the volume fraction $\phi$. Three important conclusions can be drawn from this data:
\begin{itemize}
\item The platelet thickness $\delta$, given by the slope of the dilution law, is $7.4 \pm 1.2 \text{\AA}$, much smaller than the 10~{\AA} value given in the literature \cite{Gabriel01}. This discrepancy could be due to imperfect exfoliation of the layers, leading to a lower effective concentration in the lamellar phase.
\item There is no systematic difference in $d$ and $\sigma$ between the systems formulated in pure water (at room temperature) and those with 50~\% glycerol (at 253~K). Since the phase boundaries are also the same, we conclude that the static structure of the phase is not affected by the presence of glycerol. At the same time, there is no discernible difference between the two synthesis batches.
\item The ratio $\sigma /d = 0.2 \pm 0.05$ remains almost constant as $\phi$ (and hence $d$) vary by almost an order of magnitude.
\end{itemize}

\subsection{Dynamics}\label{sec:dyn}

To investigate the dynamics of the system we consider a damped
harmonic chain, following the notations of Geisel \cite{Geisel79}.
The chain consists of $N$ particles at positions $x_j$, and we
define the displacements $u_j = x_j - x_j^0$ with respect to the
reference positions, which obey $x_{j+1}^0 - x_j^0 = d $, where $d$
is the average interparticle distance. For definiteness, we assume
periodic boundary conditions: $u_{N+1} = u_1$ and only consider odd
values for $N$.

We start by considering the case of hydrodynamically uncoupled particles, i.e.\ the energy dissipation for each particle only depends on its velocity with respect to the surrounding fluid, and not on the position or velocity of the other particles.

The equations of motion are:
\begin{subequations}
  \label{motion}
  \begin{equation}
  m \ddot{x}_j + m \gamma \dot{x}_j + \frac{\partial V}{\partial x_j} = F_j (t)
  \label{motion1}
  \end{equation}
  \begin{equation}
  V = \frac{\alpha}{2} \sum_k \left (x_{k+1} - x_k - d \right ) ^2
  \label{motion2}
  \end{equation}
 \begin{equation}
 \left \langle F_j (t) F_i (0)\right \rangle = 2 m \gamma k_B T \delta_{ij} \delta (t)
  \label{motion3}
  \end{equation}
\end{subequations}
\noindent where $m$ is the particle mass, $m\gamma$ is a damping coefficient (to be discussed in detail further on) and $F_j(t)$ is the random force acting on particle $j$ at time $t$.

The quantity of interest is the time-dependent correlation function
(depending also on the wave vector $q$):
\begin{equation}
I(q,t) = \frac{1}{N} \sum_{l,k=1}^N \left \langle
\un{e}^{i q \left ( x_l(t) - x_k(0) \right )} \right \rangle
\label{Iqt}
\end{equation}
which reduces for $t=0$ to the static structure factor (\ref{Sq}).

The treatment is quite standard: the displacements $u_j$ are
developed over the basis of normal modes (phonons). The amplitudes
and time relaxation rates of the phonons obtained from
(\ref{motion}) are used to express $I(q,t)$ by developing the
right-hand side in Eq. (\ref{Iqt}). In the following, we restrict
ourselves to the overdamped case, $\gamma \gg \sqrt{\alpha/m} = \omega_0$. In
this limit, the inertial term in Eq. (\ref{motion1}) can be
neglected and a unique relaxation rate is associated to each
phonon \cite{Note1}. The resulting
expression for $I(q,t)$ is unwieldy, so we use the cumulant
expansion \cite{Koppel72}. The first cumulant, defined as $K_1(q) \equiv - \frac{1}{I(q,0)} \left . \frac{\partial}{\partial t} I(q,t) \right | _{\, t=0}$ represents an
average relaxation rate for a given wave vector. Our experimental data is well described by a single exponential decay, $I(q,t)=I(q,0)\un{e}^{-\Omega(q)t}$, in which case the first cumulant is just the relaxation rate: $K_1(q) = \Omega(q)$, an assumption we will make throughout the analysis below.

In the uncoupled case, one has simply:
\begin{eqnarray}
\label{K1}
&&K_1(q)   = \frac{D_0}{S(q)} \, q^2\\
&&\text{with} \quad D_0=\sigma ^2 \frac{\omega_0^2}{\gamma}=\frac{k_B
T}{m \gamma}\nonumber
\end{eqnarray}

The relaxation rate obeys a diffusive law, with a diffusion constant $D(q) = D_0
/ S(q)$ which contains explicitly the well-known $1/S(q)$ dependence (``de Gennes narrowing'' \cite{deGennes59}). $D_0$ is the value of $D(q)$ at high wave vectors, where $S(q)$ reaches 1, so that one should have $D(q)\cdot S(q)/D_0 =1$ at all $q$. In other words, only the thermodynamic forces (related to $S(q)$) are taken into account, and the prefactor $D_0$ is simply proportional to the mobility of an isolated particle, $(m \gamma)^{-1}$. For a thin circular plate of radius $R$ moving normal to its plane at low Reynolds number in a fluid with viscosity $\eta$, one has \cite{Happel83}: $m \gamma = 16 \eta R$.

\begin{figure*}[htbp]
\centerline{\includegraphics[width=15cm, keepaspectratio=true]{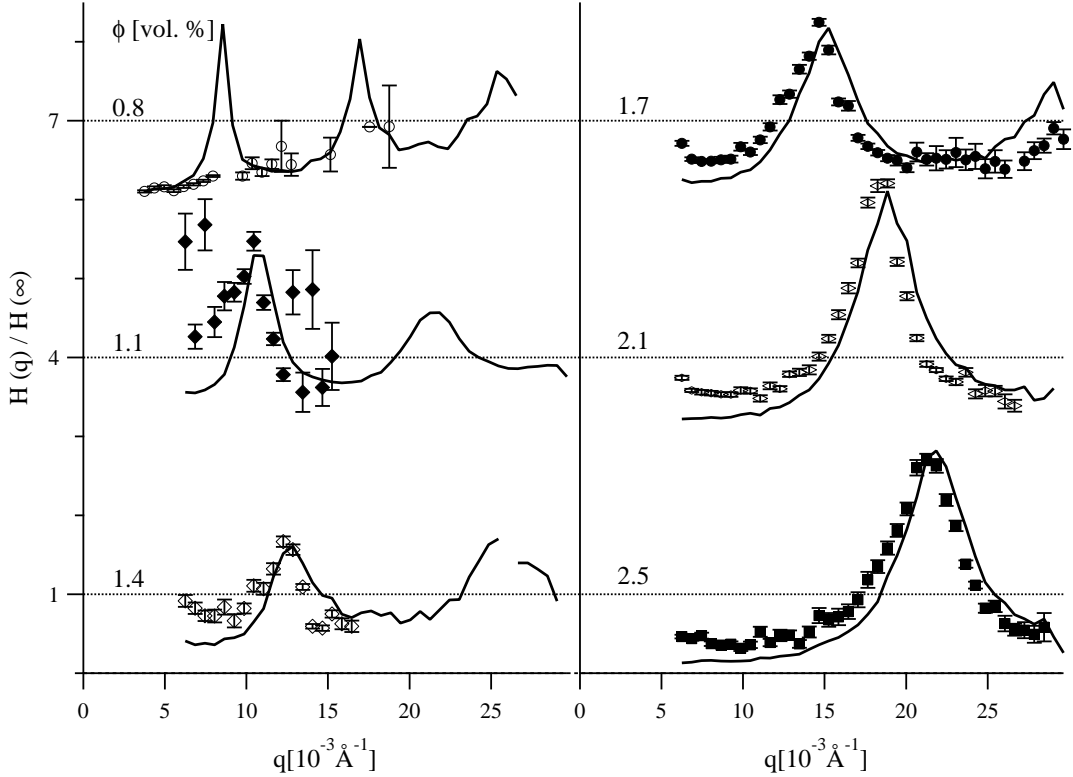}}
\caption{Experimental data for the rescaled hydrodynamic function $H(q)/H(\infty)$ (symbols) and the static structure factor $S(q)$ (solid lines) at various concentrations $\phi$ indicated alongside the curves. The left and right panel are plotted against a common $y$-axis. For clarity, the curves are shifted vertically in steps of 3 units.}
\label{Fig:Hq}
\end{figure*}

However, one must also account for the
hydrodynamic interactions, which lead to a more complex, and
generally scale-dependent, form for the dissipation. Their effect is
quantified in terms of the \textit{hydrodynamic function}, $H(q)$,
defined as \cite{Naegele97}:
\begin{equation}
H(q) = \frac{D(q)}{D_0} S(q) \label{hydro_func}
\end{equation}
In dilute solutions, $H(q) \rightarrow 1$ at high $q$ values (where
$S(q)$ saturates at 1) but is in general different from unity below
this range. We emphasize that in the simplified model discussed above, $H(q)=1$ identically (from equations (\ref{K1}) and (\ref{hydro_func}).)

Hydrodynamic interactions are essential in the dynamics of colloidal suspensions, even at moderate concentrations. In three-dimensional systems they have been studied quite extensively, both for spherical \cite{Naegele97} and for elongated \cite{Batchelor71,Shaqfeh89} particles. In recent years, the two- and one-dimensional cases have also been considered, in particular in the context of confined spherical particles \cite{Cui02,Cui04,Beatus06}. In contrast, the phase under investigation consists of very anisotropic platelets and the one-dimensional character is given by the orientational order of the phase, rather than by confinement; hence, the models cited above are probably not adapted. Since the distance between platelets along the smectic director (face to face) is much lower than their lateral extension, we only consider hydrodynamic coupling along this direction, neglecting the effect of the in-plane neighbors (edge to edge).

The simplest model that accounts for these features is the well-known Stefan equation \cite{Bird87} describing the squeezing flow between two parallel plates at low Reynolds number (in the lubrication approximation). The viscous force acting on the plates is:
\begin{equation}
F = -\frac{3 \pi \eta R^4 \dot{h}}{4 h^3} \label{Stefan}
\end{equation}
\noindent with $\eta$ the viscosity of the fluid, $R$ the radius of the (circular) plates and $h$ the gap width.

In this case, the dissipation is no longer proportional to the velocity of an individual particle with respect to the surrounding fluid, but rather to the velocity difference between neighboring particles:
\begin{eqnarray}
\label{Fvisc}
&&F_j^{\text{visc}} = - \zeta \left [ 2 \dot{x}_j  - \left ( \dot{x}_{j-1} + \dot{x}_{j+1} \right ) \right ]\\
&&\text{where} \quad \zeta = \frac{3 \pi \eta R^4}{4 d^3}\nonumber
\end{eqnarray}
For this reason, the dissipation associated to the homogeneous ($q=0$) mode vanishes: the particles can move together, at any velocity and maintaining the spacing $d$. The consequence is an infinite relaxation rate. To remove this artificial divergence, we also preserve an individual friction term (formally identical to the second term on the lhs in Equation (\ref{motion1}).) In this case, however, the quantitative value for $m \gamma$ is different from the free case, since the interaction of each particle with the fluid is ``screened'' by its neighbors. A simple yet fairly realistic model is that of a cylindrical stack of platelets: $\zeta$ corresponds then to the relative motion of particles within the stack, while $m \gamma$ is associated with a solid-like translation of the cylinder along its axis, the dissipation taking place in the surrounding medium (with an effective viscosity $\eta_{\text{eff}}$ which is in general different from that of the solvent). Using well-known formulas for the longitudinal mobility of a cylinder \cite{vanBruggen97} and considering an aspect ratio (length $L$ to diameter $2R$) of the order of 10, one has: $D_{\|} = \frac{k_B T}{\pi \eta_{\text{eff}} L}$, amounting to a friction coefficient per particle $m \gamma = \frac{k_B T}{D_{\|}} \frac{d}{L} = \pi \eta_{\text{eff}} \delta / \phi$.

With the ingredients above, and after some algebra (detailed in the Appendix) one obtains for $H(q)$ an explicit formula:
\begin{eqnarray}
\label{Hqcoupl}
H(q) &&= \frac{2}{\sqrt{1+2\beta}}{\sum_{p=0}^{N-1}} {' \atop \vphantom{\sum}} \left ( 1 - \frac{p}{N}\right ) \cos (q d p) \exp \left ( - \frac{q^2 \sigma ^2}{2} p \right ) \times \nonumber \\
&&\times z^p \, , \text{with} \, z = 1 + 1/\beta -\sqrt{1+2\beta}/\beta
\end{eqnarray}
\noindent where $N$ is the number of particles and $\beta = 2 \zeta / (m \gamma)$ is the ratio of the collective dissipation described by (\ref{Fvisc}) to the individual particle dissipation; it provides a quantitative measure of the hydrodynamic coupling. The primed sum symbol indicates that the $p=0$ term should be halved. In the high-$q$ limit, $H(q) \to H(\infty) = (2 \beta +1)^{-1/2}$ so that the corresponding diffusion constant is:
\begin{equation}
\label{Dinf}
D_{\infty} = \frac{D_0}{\sqrt{2 \beta +1}} = \frac{k_B T}{m \gamma \sqrt{4 \zeta / (m \gamma) +1}}
\end{equation}

As shown in the Appendix, for high values of the coupling constant $\beta$ the rescaled hydrodynamic function $H(q)/H(\infty)$ approaches the static structure factor $S(q)$; in other words, $D(q)$ becomes constant since the hydrodynamic effects are much stronger than the thermodynamic ones. This tendency is indeed visible in Figure \ref{Fig:Dall}, where the ``dips'' present at the peak positions for low concentrations flatten out \cite{Note2}. We would then expect that, as the concentration increases, the hydrodynamic function evolves from a constant value to the limiting shape $S(q)$. This is indeed observed in Figure \ref{Fig:Hq}, where there is a clear difference between the data at $\phi=0.8\text{~\%}$ and those at higher concentration. Indeed, at $\phi=1.1\text{~\%}$ $H(q)$ is already similar to $S(q)$ and the similarity becomes clearer above this $\phi$ value. More points within $0.8\text{~\%}\leq \phi \leq 1.1\text{~\%}$ would be needed to resolve the onset of coupling.

The similarity between $S(q)$ and $H(q)$ is very useful, since the static structure factor can be measured much more precisely and to higher $q$ values than the hydrodynamic function. We therefore rescale $H(q)$, bringing it as close as possible to $S(q)$ (the data in Figure \ref{Fig:Hq} has already been rescaled). This operation also sets the value of $D_{\infty}$ without the need of measuring $D(q)$ to very high $q$.

\begin{figure}[htbp]
\centerline{\includegraphics[width=8.4cm, keepaspectratio=true]{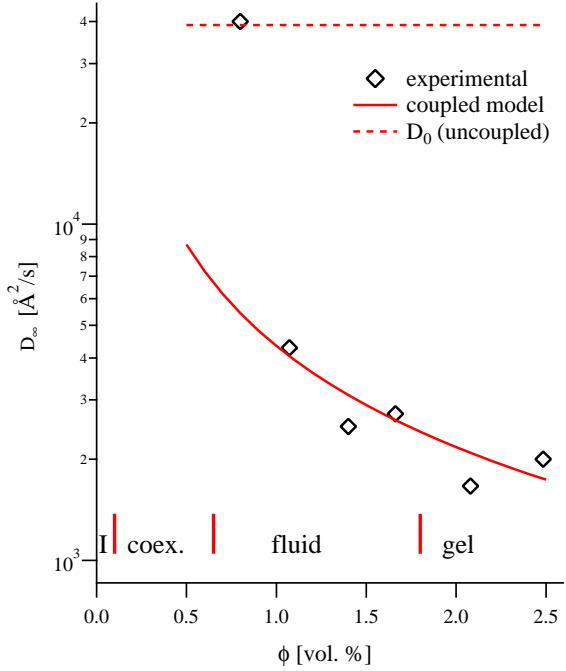}}
\caption{High-$q$ value of the diffusion constant, $D_\infty$, as a function of the volume fraction $\phi$ of platelets in a 50:50 glycerol/water mixture at 253~K. The phase diagram is also indicated: the (fluid or gel-like) lamellar phase at high concentration and an isotropic (I) phase at very low concentration, with a rather wide coexistence range.}
\label{Fig:D0}
\end{figure}

To summarize, the high-$q$ value of the diffusion coefficient in the two regimes is given by:

\begin{subequations}
  \label{Dinfeq}
  \begin{equation}
D_{\infty} = D_0 = \frac{k_B T}{16 \eta_{\text{eff}} R} \qquad \qquad \text{uncoupled}
  \label{Dinfeq1}
  \end{equation}
  \begin{equation}
D_{\infty} = \frac{k_B T}{\pi \eta_{\text{eff}} d}
\frac{1}{\sqrt{3 \left ( \frac{R}{d} \right )^4 \frac{\eta}{\eta_{\text{eff}}} +1}} \qquad \text{coupled}
  \label{Dinfeq2}
  \end{equation}
\end{subequations}
\noindent where we considered that the dissipation by squeezing flow between the particles involved the solvent viscosity $\eta = 50 \, \text{mPa~s}$ (for a 50:50 wt~\% mixture of glycerol in water at 253~K \cite{Note3}) while for the dissipation of the individual particles one needs to use the effective viscosity of the medium, $\eta_{\text{eff}}$, taken as independent of the concentration in the investigated range. The smectic repeat distance $d$ is given in Figure \ref{Fig:dilution}, so the only adjustable parameters are $\eta_{\text{eff}}$ and the platelet radius $R$. Good agreement with the experimental data is obtained with the values: $\eta_{\text{eff}} = 3.2 \eta$ and $R=3.5\, \un{\mu m}$ (see Figure \ref{Fig:D0}). Indeed, at the lowest concentration $\phi=0.8\text{~\%}$ one has $D_{\infty} = D_0$ given by Eq. (\ref{Dinfeq1}) (dashed line in Figure \ref{Fig:D0}), while at higher concentration the data is well described by the dependence (\ref{Dinfeq2}), plotted as solid line. The coupling occurs for $\phi$ between 0.8 and 1.1~\%, in agreement with the interpretation proposed above for the hydrodynamic function. It is noteworthy that no significant change in the microscopic dynamics can be detected at the transition between the fluid and the gel-like lamellar phases.

The rather large value of the platelet radius obtained above is not surprising, since the largest particles dominate both the scattering signal (with a contribution proportional to the square of the particle volume) and the correlation function, contributing the slowest relaxation. Furthermore, the coupling in Equation (\ref{Dinfeq2}) goes as the fourth power of the radius. The size distribution being very wide (see Section \ref{sec:exp}), all experimental results should be severely skewed towards the large particles.

\section{Conclusion}

We measured the static and dynamic properties of a lamellar phase composed of rigid platelets and quantified the hydrodynamic coupling between nearest neighbors (along the director of the phase). The coupling is almost absent at low concentrations, where the dissipation occurs at the level of the isolated particle, but it quickly becomes dominant at higher concentrations. The hydrodynamic function of the phase $H(q_z)$ is relatively well described by an analytical one-dimensional model.

This result is noteworthy insofar the hydrodynamic function --although indispensable for understanding the relaxation at the particle size-- is generally difficult to calculate, even for spherical colloids. To our knowledge, no explicit models have been proposed for anisotropic particles. Somewhat surprisingly, the order of the particles (which should further complicate the study) allows us in the present case to use a very simplified approach.

The high-$q$ value of the diffusion coefficient $D_{\infty}$ decreases rapidly with the concentration in the fluid lamellar phase, but it remains practically constant across the sol/gel transition and well into the gel-like regime. No sharp variation is observed at the transition, in either the dynamic or the static parameters (in particular, we do not detect any spatial inhomogeneities), showing that this transition involves longer length- and/or time-scales than probed in our experiment.

\subsection*{Acknowledgements}

We are deeply indebted to Patrick Batail, Franck Camerel, and Jean-Christophe Gabriel for providing us with the samples of {\HSb} colloidal suspensions and for helpful discussions. The ESRF is acknowledged for the provision of beam time (experiment SC-2690).

\appendix*
\section{Coupled hydrodynamics}
In this Appendix we detail the calculations of the hydrodynamic function for the harmonic model (\ref{motion}), with or without the coupling term (\ref{Fvisc}).

The boundary conditions specified in section \ref{sec:dyn} impose the eigenvector basis:
\begin{equation}
\label{eq:qn}
q_n = \frac{2 \pi}{d} \frac{n}{N} ,\, n=0, \pm 1, \ldots \pm \frac{N-1}{2}, \, \, (N \text{ odd})
\end{equation}
\noindent such that $- \pi/d < q_n < \pi/d $ (restriction to the first Brillouin zone). The normal modes (phonons) are given by $f_n (x_l) = \exp (i q_n d l) = \exp (2 i \pi l n /N)$ and the individual displacements are expressed as: $u_l(t) = \sum_{n} A_n(t) f_n(x_l)$.

The amplitude coefficients obey:

\begin{equation}
\label{eq:An}
\left \langle A_m^* A_n\right \rangle = \left \langle \left | A_n \right | ^2 \right \rangle \delta_{m,n} = \frac{k_BT}{4 N \alpha \sin ^2 (q_n d /2)} \; \delta_{m,n}.
\end{equation}

Normal mode expansion of Equation (\ref{motion1}) with the additional dissipative term (\ref{Fvisc}) yields the relaxation rate of the phonons:

\begin{eqnarray}
\label{eq:relax}
 &&\left \langle A_m^*(0) A_n (t)\right \rangle = \delta_{m,n} \left \langle \left | A_n \right | ^2 \right \rangle \exp (- \Gamma_n t), \nonumber\\
&&\text{with} \quad \Gamma_n = \frac{2 \alpha}{m \gamma} \frac{1-\cos (q_n d)}{1 + \beta \left [ 1 - \cos (q_n d) \right ]}
\end{eqnarray}
\noindent where we remind that $\beta = 2 \zeta / (m \gamma)$ is the ratio of the collective dissipation to the individual particle dissipation.

The time-dependent correlation function is then obtained by simple substitution in (\ref{Iqt}):
\begin{eqnarray}
\label{Iqtfinal}
I(q,t) &&= {\sum_{p=0}^{N-1}} {' \atop \vphantom{\sum}} \, 2 \, \frac{N-p}{N} \cos (q d p) \exp \left [ - \frac{q^2 \sigma ^2}{2} \times \right . \nonumber \\
&&\times \left . \frac{1}{N} \sum_{n} \frac{1- \cos (q_n d p) \exp (- \Gamma_n t)}{1 - \cos (q_n d)} \right ]
\end{eqnarray}
\noindent and the first cumulant is obtained according to (\ref{K1}), yielding for $H(q)$:

\begin{eqnarray}
\label{Hqcalcul}
H(q) &&= {\sum_{p=0}^{N-1}} {' \atop \vphantom{\sum}} \, 2 \, \frac{N-p}{N} \cos (q d p) \exp \left ( - \frac{q^2 \sigma ^2}{2} p \right ) \times \nonumber \\
&&\times C(p,\beta)
\end{eqnarray}

\noindent where the coefficients $C(p,\beta) = \frac{1}{N} \sum_{n} \frac{\cos (q_n d p) }{1 + \beta \left [ 1 - \cos (q_n d) \right ]}$ can be reduced (e.g.\ by going to the continuum limit and performing a complex integral over the unit circle) to: $C(p,\beta) = \frac{z^p}{\sqrt{2\beta + 1}}$, finally yielding equation (\ref{Hqcoupl}) above. For the uncoupled case $\beta = 0$ this result is greatly simplified, since the sum over the modes in Equation (\ref{Hqcoupl}) becomes: $\frac{1}{N} \sum_{n} \cos (q_n d p) = \delta_{p,0}$ and thus $H(q,\beta=0) = 1$, leading to the second equality in Equation (\ref{K1}).

The strongly coupled form $\displaystyle \lim_{\beta \rightarrow \infty} H(q) = \frac{S(q)}{\sqrt{2\beta + 1}}$ is obtained by noting that in this limit $z$ approaches 1 from below, so that the damping factor $\exp \left [ - (\frac{q^2 \sigma ^2}{2} - \ln z) p \right ]$ is dominated by the first term in the exponent. One can therefore drop the $z^p$ factor in Eq. (\ref{Hqcalcul}) and retrieve an obvious expansion of the static structure factor $S(q)$. Of course, for this to hold the scattering vector must also be above a certain value, otherwise the first term in the exponent might be too small. We checked that for reasonable numerical values the condition is fulfilled for all accessible values of $q$.

\end{document}